%% file: main.tex
\pdfoutput=1

\documentclass[11pt]{article}

\usepackage[final]{acl}

\usepackage{times}
\usepackage{latexsym}
\usepackage{enumitem}
\usepackage{graphicx} 
\usepackage{bm}
\usepackage{amsmath}
\usepackage{multirow}
\usepackage{subfigure} 
\newcommand{\ie}{\emph{i.e., }}
\newcommand{\eg}{\emph{e.g., }}
\usepackage{tabularx}

\usepackage[T1]{fontenc}

\usepackage[utf8]{inputenc}

\usepackage{microtype}

\usepackage{inconsolata}

\title{Text-like Encoding of Collaborative Information in Large Language Models for Recommendation}

\author{
 \textbf{Yang Zhang\textsuperscript{1}},
 \textbf{Keqin Bao\textsuperscript{1}},
 \textbf{Ming Yan\textsuperscript{1}}, 
 \textbf{Wenjie Wang\textsuperscript{2}}, \\ 
 \textbf{Fuli Feng\textsuperscript{1*}},
 \textbf{Xiangnan He\textsuperscript{1*}}
\\
 \textsuperscript{1}University of Science and Technology of China, \\
 \textsuperscript{2}National University of Singapore
\\
 \small
 zyang1580@gmail.com, \{baokq, ym689\}@mail.ustc.edu.cn, \\
 \small
 \{wenjiewang96,fulifeng93,xiangnanhe\}@gmail.com
}

\newcommand \footnoteONLYtext[1]
{
	\let \mybackup \thefootnote
	\let \thefootnote \relax
	\footnotetext{#1}
	\let \thefootnote \mybackup
	\let \mybackup \imareallyundefinedcommand
}

\begin{document}
\maketitle

\footnoteONLYtext{*Corresponding authors.}

\begin{abstract}

When adapting Large Language Models for Recommendation (LLMRec), it is crucial to integrate collaborative information. Existing methods achieve this by learning collaborative embeddings in LLMs' latent space from scratch or by mapping from external models. However, they fail to represent the information in a text-like format, which may not align optimally with LLMs. To bridge this gap, we introduce BinLLM, a novel LLMRec method that seamlessly integrates collaborative information through text-like encoding. BinLLM converts collaborative embeddings from external models into binary sequences --- a specific text format that LLMs can understand and operate on directly, facilitating the direct usage of collaborative information in text-like format by LLMs. Additionally, BinLLM provides options to compress the binary sequence using dot-decimal notation to avoid excessively long lengths. Extensive experiments validate that BinLLM introduces collaborative information in a manner better aligned with LLMs, resulting in enhanced performance. We release our code at \url{https://github.com/zyang1580/BinLLM}. 
\end{abstract}

\input{latex/1_intro}

\input{latex/3_method}

\input{latex/4_exp}

\input{latex/5_related}

\input{latex/6_conclusion}
\input{latex/7_Appendenx}
\bibliography{custom,ref}

\appendix



\end{document}

%% file: latex/1_intro.tex
\section{Introduction}


Due to the remarkable power of large language models (LLMs), there is a growing focus on adapting them for recommender systems (LLMRec), which has seen significant progress in the past year~\citep{tallrec,bigrec,LLMRec-survey,harte2023leveraging,tiger,wei2023llmrec}. In recommendation, collaborative information, which delineates the co-occurrence patterns among user-item interactions, has emerged as a pivotal component in modeling user interests, especially for active users and items~\cite{collm}. However, this information exists in a different modality from textual data and thus presents a challenge in directly leveraged by LLMs like textual information~\citep{collm,ctrl,bigrec}. To enhance recommendation quality, it is undoubtedly crucial to seamlessly integrate collaborative information into LLMs.

To date, two integration strategies have emerged. The first strategy resembles latent factor models~\cite{koren2009matrix} by incorporating additional tokens and corresponding embeddings into LLMs to represent users and items, subsequently fitting interaction data to implicitly capture collaborative information within the embeddings~\citep{RUCidcf,yongfengID}. However, this approach suffers from low learning efficacy due to the inherent low-rank nature of the information, leading to tokenization redundancy within LLMs~\cite{LLMCompression,collm}. To address these challenges, an alternative approach leverages an external latent factor model to capture the information, which is then mapped into the LLM token embedding space~\citep{collm,e4srec,llara}, circumventing the need to learn it from scratch. While effective, this method introduces the additional overhead of training the mapping model.

Whether learning collaborative information directly from scratch in the LLM token embedding space or mapping it from external models, the resulting representations diverge significantly from the LLM's original textual-level encoding. This, to a certain extent, hampers the full utilization of LLMs' capabilities, as LLMs are initially trained on textual data and excel at processing textually encoded information. For instance, introducing new tokens alters the generative space of LLMs, potentially compromising their original functionalities, let alone capitalizing on their capabilities. Therefore, exploring text-like encoding of collaborative information in LLMs holds immense promise. Nevertheless, it poses challenges due to the inherent differences between textual and collaborative information modalities~\citep{collm}.





In this study, we delve into the central theme of encoding collaborative information in LLMs for recommendation, an area of promise yet not explored in LLMRec. 
The crux lies in transforming collaborative information into a sequence formatted like text.
We believe this text-like sequence need not be comprehensible to humans; rather, it should be interpretable by LLMs for effective utilization, such as facilitating reasoning tasks like discerning user and item similarities through sequence comparisons. Thus, this text sequence does not necessarily have to adhere to conventional natural language patterns.


To this end, we introduce \textit{BinLLM}, an innovative LLMRec approach that integrates collaborative information into LLMs using a text-like encoding strategy. We transform the collaborative embeddings obtained from external models into binary sequences, treating them as textual features directly usable by LLMs. This design is motivated by two primary considerations: 1) the feasibility of binarizing collaborative embeddings without compromising performance~\citep{hashGNN}; 2) LLMs can naturally perform bitwise operations or do so after instruction tuning~\citep{bitwiseinLLM}, enabling the comparison of similarities between binarized sequences. Taking a step further, we explore representing the binary sequence in dot-decimal notation~\citep{dot-notation}, resulting in shorter representations, akin to converting binary sequences to IPv4 addresses. By fine-tuning LLMs with recommendation instruction data containing such encoded collaborative information, we could leverage both textual semantics and collaborative data for recommendation without modifying the LLMs.

The main contributions of this work are summarized as follows:
\begin{itemize}[leftmargin=*, itemsep=0pt,parsep=2pt]

    \item We emphasize the significance of text-like encoding for collaborative information in LLMRec to enhance alignment with LLMs.

\item We introduce \textit{BinLLM}, a novel method that efficiently encodes collaborative information textually for LLMs by converting collaborative embeddings into binary sequences.
    
\item We perform comprehensive experiments on two datasets, showcasing the effectiveness of our approach through extensive results.
\end{itemize}

%% file: latex/3_method.tex
\section{Methodology}
In this section, we introduce our BinLLM method, starting with presenting the model architecture and followed by a description of the tuning method.

\subsection{Model Architecture}
\begin{figure*}
    \centering
    \includegraphics[width=0.95\textwidth]{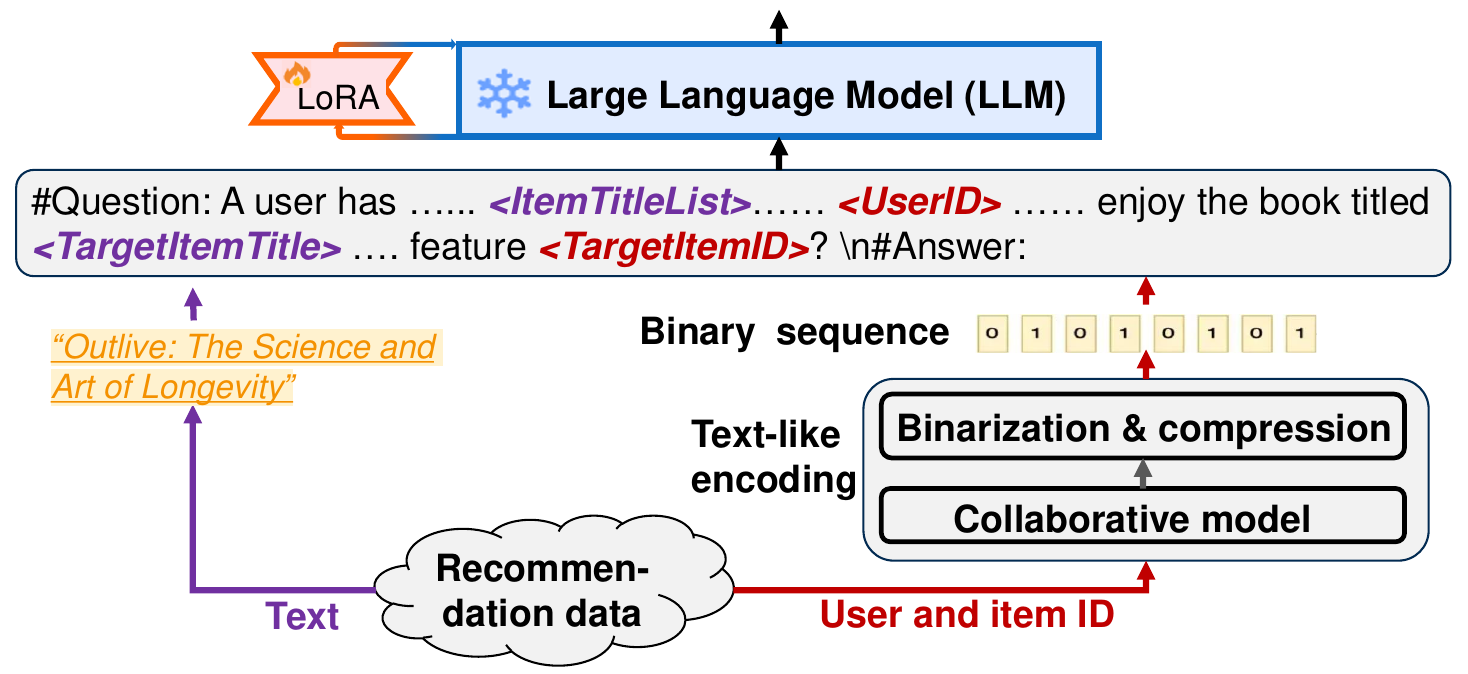}
    \caption{Model architecture overview of our BinLLM. The purple line is used to fill the text fields in the prompt template, introducing textual information like item titles, while the red line is used to fill the ID fields in the prompt template, introducing collaborative information.}
    \label{fig:binLLM}
\end{figure*}

Figure~\ref{fig:binLLM} depicts the model architecture of BinLLM, comprising two main components: prompt generation and LLM prediction. Similar to previous approaches, we convert recommendation data into prompts and then input them directly into LLMs for prediction. However, the key distinction of BinLLM is that it represents collaborative information in a text-like format by converting collaborative embeddings into binary sequences. We next delve into the specifics of these two components.

\subsubsection{Prompt Construction}\label{sec:prompt}
As depicted in Figure~\ref{fig:binLLM}, we construct prompts using a template featuring empty fields, encompassing both textual fields (\eg ``<ItemTitleList>") and ID fields (\eg ``<UserID>"). See the template example in Table~\ref{tab:prompt-example}. By populating these fields with corresponding users' data, we can generate personalized prompts for recommendation purposes. The textual fields are utilized to incorporate textual information, which can be directly filled with corresponding textual data from the recommendation dataset, such as historical item titles in the "<ItemTitleList>" fields. The ID fields are designated for embedding collaborative information, which is acquired through a Text-like Encoding (TE) module.
Next, we delve into the encoding process of collaborative information.

\begin{table}
\centering
  \caption{
        Example of the used prompt template, using the same format as CoLLM.
    }
 \label{tab:prompt-example}
    \begin{tabular}{p{7.5cm}}  
    \hline
        \#Question: A user has given high ratings to the following books: <ItemTitleList>. Additionally, we have information about the user's preferences encoded in the feature <UserID>. Using all available information, make a prediction about whether the user would enjoy the book titled <TargetItemTitle> with the feature <TargetItemID>? Answer with "Yes" or "No". \textbackslash n\#Answer:\\
    \hline
    \end{tabular}
\end{table}

\par \textbf{\textit{Text-like Encoding of Collaborative Information.}} 
To better integrate with LLMs, we aim to encode collaborative information in a text-like format. To accomplish this, we convert collaborative information into a binary sequence, enabling LLMs to perform bitwise operations for reasoning. The encoding model involves two components: 
1) Collaborative Model, a conventional latent factor module capable of encoding collaborative information as numerical latent vectors (\ie collaborative embeddings).
2) Binarization \& Compression Module, utilized to transform collaborative embeddings into binary sequences or further compressed formats.

\noindent $\bullet$ \textbf{Collaborative model.} 
Given a user $u$ and an item $i$, the collaborative model generates corresponding embeddings for them, denoted as $\bm{e}_{u}$ and $\bm{e}_{i}$, respectively. Formally,
\begin{equation}
 \begin{split}
      &\bm{e}_{u} = f_c(u;\theta)\\ &\bm{e}_{i}=f_c(i;\theta),
 \end{split}
\end{equation}
where $f_c$ represents the collaborative model parameterized by $\theta$. Here, $\bm{e}_{u} \in \mathcal{R}^d $ and $\bm{e}_{i}\in \mathcal{R}^d$ are $d$-dimensional embeddings that encode collaborative information for the user and item, respectively.

\noindent $\bullet$ \textbf{Binarization \& compression.} After obtaining the collaborative embeddings, this component is used to convert them into binary sequences, with the option to compress the sequences. 

\textit{Binarization}. 
To binarize the collaborative embeddings, we generally follow the mechanism proposed by \citet{hashGNN}. Firstly, we transform the collaborative embeddings into a suitable space using a fully connected layer and then apply the sign function to obtain the binary results. Formally, for collaborative embeddings $\bm{e}_{u}$ and $\bm{e}_{i}$ of user $u$ and item $i$, they are converted into binary sequences as follows:
\begin{equation}\label{eq:hash_emb}
\begin{split}
    & \bm{h}_u = sign(\sigma(W\bm{e}_u + b))\, \\
    & \bm{h}_i = sign(\sigma(W\bm{e}_i + b)),
    \end{split}
\end{equation}
where $\bm{h}_u \in \{0,1\}^d$ and $\bm{h}_i\in \{0,1\}^d$ denote the obtained binary representation of collaborative information for the user and item, respectively. Here, $W \in \mathcal{R}^{d\times d}$ and $b\in \mathcal{R}^d$ are the weights and bias for the fully connected layer, $\sigma(\cdot)$ represents the \textit{tanh} activation function, and $\text{sign}(\cdot)$ denotes the sign function. For a numerical value $x$, we have:
\begin{equation}
    sign(x) = \begin{cases}
        1, \quad \text{if $x>0$}\\
        0, \quad \text{else}
    \end{cases}.
\end{equation}


Through this method, we convert the numerical collaborative embeddings into binary sequences (e.g., '010110....'). These sequences can be directly inputted into LLMs and utilized for operations such as computing logical 'AND', thereby aiding in user preference reasoning.

\textit{Compression}. A limitation of binary sequences is their relatively long length, which poses a challenge for LLMs not proficient in handling lengthy sequences. Moreover, long sequences can constrain the inference efficiency of LLMRec. We thus consider compressing the binary sequences while keeping them leverageable by LLMs. Given that IPv4~\cite{ipv4} is originally encoded from binary sequences and the Web includes sufficient knowledge about IPv4, the LLMs trained on the Web data could potentially understand the dot-decimal notation used by IPv4. Therefore, we consider compressing the binary embeddings in dot-decimal notations~\cite{dot-notation}. We convert every eight binary digits into a decimal number, ranging from 0 to 255, and use the full stop (dot) as a separation character. Here is an example of compressing a 32-bit binary sequence:
\begin{equation}\label{eq:ip}
\underbrace{10101100}_{172.}\underbrace{00010000}_{16.}\underbrace{11111110}_{254.}\underbrace{00000001}_{1}.
\end{equation}
Here, ``172.16.254.1" is the compressed result, which significantly reduces the representation length. Notably, the compression is optional, and its usage depends on the length of the original binary sequence.

\subsubsection{LLM Prediction} 
Once the empty fields in the prompt template are filled, the resulting prompt is fed into the LLMs for prediction. Similar to prior research~\cite{tallrec,collm}, given the absence of specific recommendation pre-training in LLMs, we introduce an additional LoRA module~\cite{lora} for recommendation prediction. Formally, for a generated prompt $p$, the prediction can be formulated as:
\begin{equation}
    \hat{y} = LLM_{\hat{\Phi}+\Phi^{'}}(p),
\end{equation}
where $\hat{\Phi}$ represents the pre-trained LLM's parameters, $\Phi^{'}$ denotes the LoRA model parameters, and $\hat{y}$ represents the prediction results, which could be the predicted next item or the predicted likelihood of liking a candidate item, depending on the task.

\subsection{Training} 
In our model architecture, two modules require training: the text-like encoding module and the LoRA module. The tuning for the text-like encoding module focuses on learning to generate the binary sequence for collaborative information, independent of the LLMs. The tuning for LoRA aims to instruct the LLM in making recommendations by leveraging collaborative information. We now present the two tuning paradigms, respectively.

\subsubsection{Pre-training for Text-like Encoding}

To train the text-like encoding module, we directly utilize the binarized representation from Equation~\eqref{eq:hash_emb} to fit the training data. Formally, let $\mathcal{D}$ denote the training data, and $(u,i,t) \in \mathcal{D}$ denote an interaction between user $u$ and item $i$ with label $t$. We train the module by minimizing the following optimization problem:
\begin{equation}
    \mathop{minimize} \limits _{\theta, W,b} \sum_{(u,i,t)\in \mathcal{D}}{ \ell(t,\bm{h}_u^{\top}\bm{h}_i) }, 
\end{equation}
where $\{\theta, W, b\}$ denote the model parameters in our text-like encoding module as discussed in Section~\ref{sec:prompt}, $\bm{h}_u$ and $\bm{h}_i$ denote the binary representations\footnote{Notably, during training, we will convert the binary values of 0 to -1 for $\bm{h}_u$ and $\bm{h}_i$ following the approach used in prior work~\cite{hashGNN}. } obtained from Equation~\eqref{eq:hash_emb}, $\bm{h}_u^\top \bm{h}_i$ represents the predicted likelihood of user $u$ liking item $i$, and $\ell(\cdot)$ denotes the common recommendation loss, in this work, the binary cross-entropy loss.


Notably, the sign function lacks smoothness, and its gradient is ill-defined as zero, posing an apparent challenge for back-propagation. To enable training the model in an end-to-end fashion, we approximate the gradient using the straight-through estimator (STE), following the approach outlined by \citet{hashGNN}. That is, we directly use the gradients of the output as the gradients of the input for the sign function.

\subsubsection{LoRA Tuning}
To tune the LoRA module, we consider two tuning methods: intuitive tuning and two-step tuning. 

\textbf{Intuitive tuning:} This method directly tunes the LoRA module from scratch with the prompts that contain the collaborative information. 

\textbf{Two-step tuning:} 
In intuitive tuning, a potential challenge arises in scenarios like rating prediction tasks, where binary representations can serve as highly effective features with relatively low learning complexity\footnote{Because the model could achieve satisfactory results by solely performing bitwise "AND" operations on the collaborative representations of the given user and candidate item, referencing the learning process of binary representation.}. Incorporating collaborative information from scratch might cause the model to overly depend on these features, potentially neglecting other attributes akin to learning shortcut features. To address this, we propose an additional two-step tuning strategy. Initially, we train the model using a prompt that excludes collaborative information. Subsequently, we refine the model further by fine-tuning it using the complete prompt that contains the collaborative information.


%% file: latex/4_exp.tex
\section{Experiments}
In this section, we conduct experiments to answer the following research questions: 

\noindent\textbf{RQ1}: 
Does BinLLM effectively incorporate collaborative information into LLMs to improve recommendation performance? How does its performance compare with that of existing methods? 

\noindent\textbf{RQ2}: How do our design choices influence the performance of the proposed method BinLLM?

\subsection{Experimental Settings}

\paragraph{Recommendation Task.} 
Given that this is an initial exploration of text-like encoding for collaborative information, our experiments primarily concentrate on the click/rating prediction task, with other recommendation tasks being ignored. Specifically, we aim to predict whether a user $u$ (comprising other profile information such as historical interactions) would click on/like a given candidate item $i$. The task aligns with that of CoLLM, which investigates the utilization of collaborative information for recommendation through embedding mapping in latent space. Hence, our experimental setup generally follows that of CoLLM.

\begin{table}[t]
\caption{Statistics of the processed datasets.}
\label{tab:dataStatics}
\renewcommand\arraystretch{0.9}
\resizebox{0.49\textwidth}{!}{
\begin{tabular}{cccccc}
\hline
Dataset&\#Train&\#Valid&\#Test&\#User&\#Item
\\ \hline
ML-1M&33,891&10,401&7,331&839&3,256 \\
Amazon-Book&727,468&25,747&25,747&22,967&34,154\\\hline

\end{tabular}
}
\end{table}

\paragraph{Datasets.} We conduct experiments on two representative datasets:
\begin{itemize}[leftmargin=*, itemsep=0pt,parsep=2pt]
    
    \item \textbf{ML-1M}~\citep{movielens}: This refers to a widely recognized movie recommendation benchmark dataset, MovieLens-1M\footnote{\url{https://grouplens.org/datasets/movielens/1m/}}, provided by GroupLens research. The dataset comprises user ratings for movies and includes textual information for users and items, such as movie titles.

    \item \textbf{Amazon-Book}~\citep{amazondata}: This pertains to the "Books" subset within the renowned Amazon Product Review dataset\footnote{\url{https://nijianmo.github.io/amazon/index.html}}. This dataset aggregates user reviews of books from Amazon, encompassing both the review score and review comments. Additionally, it includes textual information about the items.
\end{itemize}
For dataset processing, we adhere entirely to the setup of CoLLM, encompassing label processing and data selection/splitting methods. The statistics of the processed datasets are presented in Table~\ref{tab:dataStatics}.

\begin{table*}[th]
\caption{Overall performance comparison on the ML-1M and Amazon-Book datasets. ``Collab.'' denotes collaborative recommendation methods. ``Rel. Imp.'' denotes the relative improvement of BinLLM compared to baselines, averaged over the two metrics.}
\centering
\begin{tabular}{cc|ccc|ccc}
\hline
\multicolumn{2}{c|}{Dataset}                                       & \multicolumn{3}{c|}{ML-1M}  & \multicolumn{3}{c}{Amazon-Book} \\ \hline
\multicolumn{2}{c|}{Methods}                                       & AUC    & UAUC   & Rel. Imp. & AUC      & UAUC    & Rel. Imp.  \\ \hline
\multicolumn{1}{c|}{\multirow{4}{*}{Collab.}} & MF                 & 0.6482 & 0.6361 & 12.9\%    & 0.7134   & 0.5565  & 14.7\%     \\
\multicolumn{1}{c|}{}                         & LightGCN           & 0.5959 & 0.6499 & 15.8\%    & 0.7103   & 0.5639  & 14.2\%     \\
\multicolumn{1}{c|}{}                         & SASRec             & 0.7078 & 0.6884 & 3.0\%     & 0.6887   & 0.5714  & 15.3\%      \\
\multicolumn{1}{c|}{}                         & DIN                & 0.7166 & 0.6459 & 5.6\%     & 0.8163   & 0.6145  & 2.0\%      \\ \hline
\multicolumn{1}{c|}{LM+Collab.}               & CTRL (DIN)         & 0.7159 & 0.6492 & 5.4\%     & 0.8202   & 0.5996  & 3.0\%      \\ \hline
\multicolumn{1}{c|}{\multirow{3}{*}{LLMRec}}  & ICL                & 0.5320 & 0.5268 & 35.8\%    & 0.4820   & 0.4856  & 50.7\%     \\
\multicolumn{1}{c|}{}                         & Prompt4NR & 0.7071 & 0.6739 & 4.1\%     & 0.7224   & 0.5881  & 10.9\%     \\
\multicolumn{1}{c|}{}                         & TALLRec            & 0.7097 & 0.6818 & 3.3\%     & 0.7375   & 0.5983  & 8.2\%      \\ \hline
\multicolumn{1}{c|}{}                         & PersonPrompt             & 0.7214 & 0.6563 &  4.5\%          & 0.7273   & 0.5956  &    9.9\%        \\
\multicolumn{1}{c|}{LLMRec+Collab.}              & CoLLM-MF           & 0.7295 & 0.6875 & 1.5\%         & 0.8109   & 0.6225  & 1.7\%          \\
\multicolumn{1}{c|}{}                         & CoLLM-DIN          & 0.7243 & 0.6897 & 1.7\%         & 0.8245   & \textbf{0.6474}  & -1.0\%          \\ \hline
\multicolumn{1}{c|}{Ours}                     & BinLLM             &    \textbf{0.7425}    & \textbf{0.6956}       &    -       &    \textbf{0.8264}      & 0.6319        &  -          \\ \hline
\end{tabular}
\label{tab-main}
\end{table*}

\paragraph{Compared Methods.} 
In this work, we implement BinLLM with Matrix Factorization~\citep{koren2009matrix} as the collaborative model in its text-encoding module. To assess the effectiveness of BinLLM, we compare it with four categories of methods: conventional collaborative filtering methods (MF, LightGCN, SASRec, DIN), LLMRec methods without integrating collaborative information (ICL, Prompt4NR, TALLRec), LLMRec methods with integrated collaborative information (PersonPrompt, CoLLM), and methods combining language models and collaborative models (CTRL).

\begin{itemize}[leftmargin=*,itemsep=0pt,parsep=3pt]
    \item \textbf{MF~\citep{koren2009matrix}}: This refers to a classic latent factor-based collaborative filtering method --- Matrix Factorization.

     \item \textbf{LightGCN~\cite{lightgcn}}: This is one representative graph-based collaborative filtering method, utilizing graph neural networks to enhance collaborative information modeling.
     
      \item \textbf{SASRec~\cite{SASRec}}: 
     This is a representative sequential-based collaborative filtering method that utilizes self-attention for modeling user preferences.
      

      \item \textbf{DIN~\cite{DIN}}: 
      This is a representative collaborative Click-Through Rate (CTR) model, which employs target-aware attention to activate the most relevant user behaviors, thereby enhancing user interest modeling.
      

      \item \textbf{CTRL (DIN)~\cite{ctrl}}: 
      This is a state-of-the-art (SOTA) method for combining language and collaborative models through knowledge distillation. We implement its collaborative model as DIN.
      
      
     \item \textbf{ICL~\cite{ICL}}: 
     This is an In-Context Learning-based LLMRec method, which directly asks the original LLM for recommendations.

     \item \textbf{Prompt4NR~\cite{prompt4nr}}: 
     This is a state-of-the-art (SOTA) soft prompt tuning-based LLMRec method. Initially designed to leverage the language model (LM), we extend it to utilize LLMs, taking the implementation in CoLLM~\cite{collm}.
     

     \item \textbf{TALLRec~\cite{tallrec}}: 
     This is a state-of-the-art LLMRec method that aligns LLMs with recommendations through instruction tuning.

     \item \textbf{PersonPrompt~\cite{personprompt}}: This is a LLMRec method, which integrates collaborative information by adding new tokens and token embeddings to represent users and items. It could be regarded as a personalized soft-prompt tuning method.
     

     \item \textbf{CoLLM~\cite{collm}}: 
     This is a state-of-the-art LLMRec method that integrates collaborative information by mapping collaborative embeddings into the latent space of the LLM. We consider two implementations: CoLLM-MF, which utilizes MF to extract collaborative embeddings, and CoLLM-DIN, which uses the DIN to extract collaborative embeddings.
\end{itemize}


\paragraph{Hyper-parameters and Evaluation Metrics.} For all methods, we strictly adhere to the hyperparameter settings outlined in the CoLLM paper~\citep{collm}, with Vicuna-7B used as the employed LLM. It's worth noting that for our method, we set the dimension of the collaborative embeddings (\textit{i.e.}, the length of the binary representations in Equation~\eqref{eq:hash_emb}) to 32 by default. Considering the length is not very large, we choose not to perform compression in our text-like encoding module by default. We tune the hyper-parameters based on the AUC metric on the validation dataset.

Regarding evaluation metrics, we employ two widely used metrics for click/rating prediction: AUC (Area under the ROC Curve), which measures the overall prediction accuracy, and UAUC (AUC averaged over users), which provides insights into the ranking quality for users.


\begin{figure}[t]
\centering
\subfigure[\textbf{ML-1M Warm}]{\includegraphics[width=1.475in,height=1.15in]{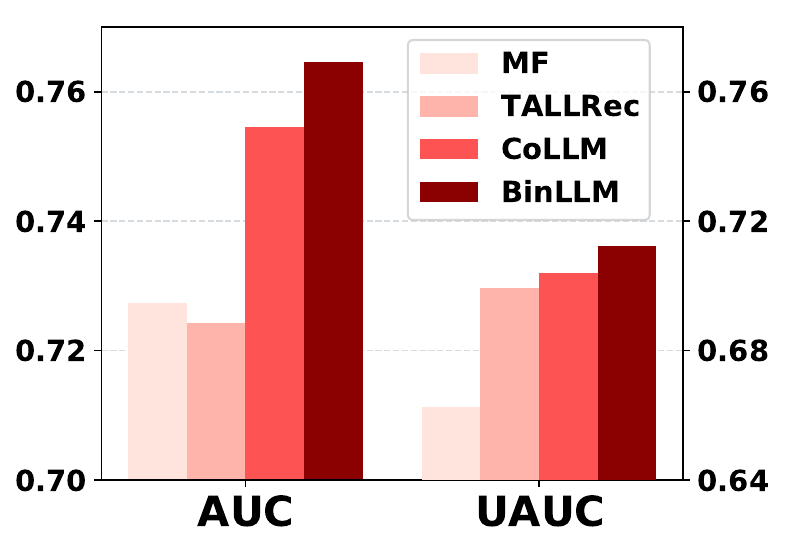}}
\subfigure[ \textbf{Amazon-book Warm}]{ \includegraphics[width=1.475in,height=1.15in]{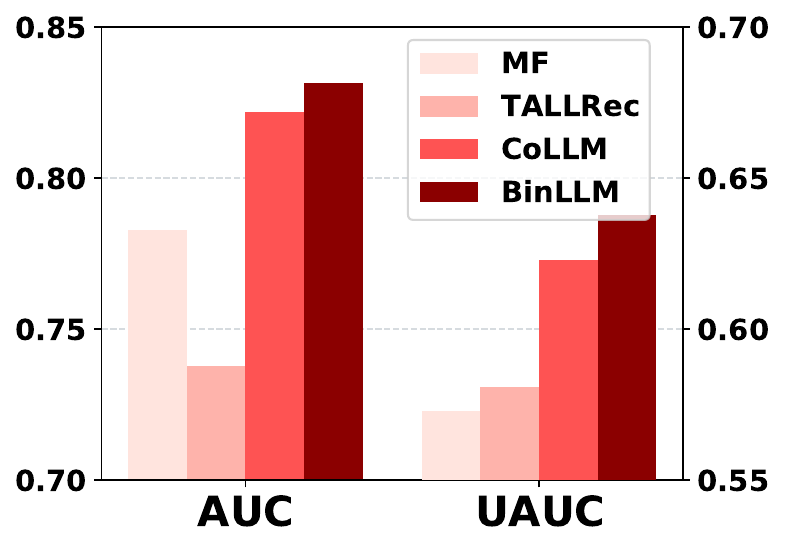}}
\subfigure[ \textbf{ML-1M Cold}]{ \includegraphics[width=1.46in,height=1.15in]{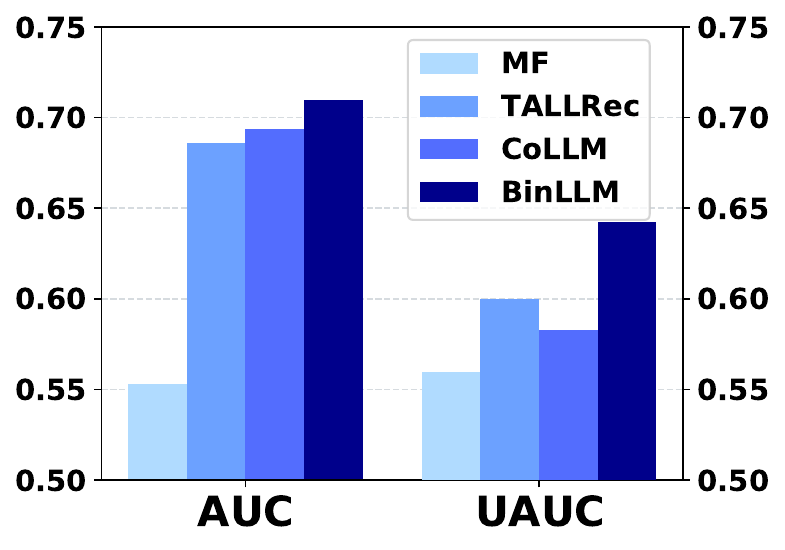}}
\subfigure[ \textbf{Amazon-book Cold}]{ \includegraphics[width=1.46in,height=1.15in]{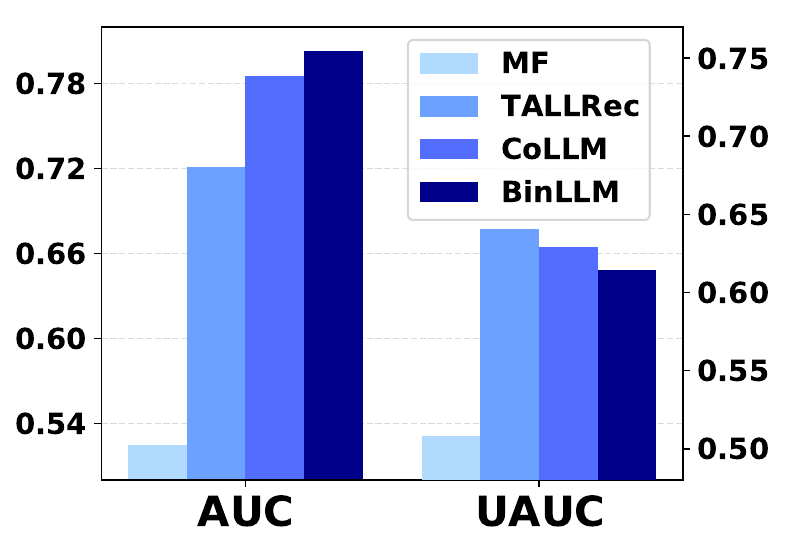}}
\caption{Performance comparison in warm and cold scenarios on ML-1M and Amazon-Book. The left y-axis represents AUC, while the right one represents UAUC.}
\label{fig:warm-cold}
\end{figure}

\subsection{Performance Comparison} 

In this subsection, we initially examine the overall performance of the compared methods and subsequently analyze their performance in warm-start and cold-start scenarios, respectively.

\subsubsection{Overall Performance (RQ1)}
We summarize the overall performance of the compared methods in Table~\ref{tab-main}. From the table, we draw the following observations:
\begin{itemize}[leftmargin=*,itemsep=0pt,parsep=3pt]
\item When compared to baselines, our BinLLM achieves the best performance overall, except when compared to CoLLM-DIN on the UAUC metric. 
These results confirm the superiority of BinLLM in leveraging both collaborative information and the power of LLMs to achieve better recommendation performance.

\item 
Comparing LLMRec methods that integrate collaborative information with LLMRec methods that do not consider collaborative information, we observe that incorporating collaborative information generally improves performance and enables LLMRec to surpass traditional collaborative and LM-based methods. These results underscore the importance of integrating collaborative information into LLMs for recommendation.

\item Comparing BinLLM with existing LLMRec methods that also consider collaborative information, our BinLLM consistently outperforms CoLLM-MF and PersonPrompt. Compared with CoLLM-DIN, BinLLM still achieves better results except for the UAUC metric on Amazon-book. Considering that CoLLM-DIN employs a more advanced collaborative model while BinLLM relies solely on MF, these results confirm that encoding collaborative information in a text-like manner better aligns with LLMs, allowing us to leverage their power for recommendation more effectively.

\item 
Among LLMRec methods that consider collaborative information, PersonPrompt, which learns token embeddings for users and items from scratch, performs the worst, significantly lagging behind others. This can be attributed to the low learning efficacy resulting from the introduction of additional tokens and token embeddings.
\end{itemize}

\subsubsection{Warm and Cold Performance}
When integrating collaborative information into LLMRec, one consideration is to enhance their warm-start performance, enabling them to achieve good performance in both warm-start and cold-start scenarios. We now investigate the performance in the two scenarios. Specifically, we adhere to the protocol outlined in the CoLLM paper~\cite{collm} to partition the testing data into warm data and cold data based on the interaction count of users and items, and subsequently evaluate the model on them. We summarize the results in Figure~\ref{fig:warm-cold}. Here, we compare four representative methods: MF, TALLRec, CoLLM-MF, and BinLLM.

According to the figure, in the warm scenarios, TALLRec, an LLMRec method without considering collaborative information, performs worse than MF, while both CoLLM and BinLLM outperform MF, with BinLLM being the best. These results indicate that collaborative information is important for warm-start performance, and our text-like encoding has superiority in combining the information with LLMs. In the cold-start scenarios, all LLMRec methods outperform MF, confirming the superiority of LLMRec in cold-start scenarios. Moreover, BinLLM enhances the cold-start performance compared to CoLLM in most cases, possibly due to the binarized embeddings having better generalization.


\subsection{In-depth Analyses (RQ2)} 
In this subsection, we conduct experiments to analyze the influence of BinLLM's different components on its effectiveness.

\begin{table}[]
\caption{Results of the ablation studies on ML-1M and Amazon-Book, where ``TO", ``IO", ``IT" denote ``Text-Only", ``ID-Only", ``Intuitive-Tuning", respectively.  }
\resizebox{0.49\textwidth}{!}{
\begin{tabular}{c|cc|cc}
\hline
   Datasets             & \multicolumn{2}{c|}{ML-1M} & \multicolumn{2}{c}{Amazon-book} \\ \hline
Methods         & AUC          & UAUC        & AUC            & UAUC           \\ \hline
BinMF           & 0.7189       & 0.6654      & 0.8087         & 0.5895         \\
BinLLM-TO & 0.7097       & 0.6818      & 0.7375         & 0.5983         \\
BinLLM-IO   &      0.7307        & 0.6797            &  0.8173              &   0.5919             \\
BinLLM-IT       &      0.7286        &    0.6842         &     0.8246           &      0.6165          \\ \hline
BinLLM          & 0.7425       & 0.6956       & 0.8264         & 0.6319         \\ \hline
\end{tabular}
}
\label{tab-ablation}
\end{table}

\subsubsection{Ablation Study}
We first further verify the benefits of introducing text-like encoding of collaborative information into LLMs. Specifically, we compare the default BinLLM with the following variants: 1) BinMF, which avoids using the LLM but directly utilizes the binary representations for recommendations like MF, 2) BinLLM-TO, which removes the ID field from BinLLM's prompt template, \textit{i.e.,} only using the text information, 3) BinLLM-IO, which removes the text field from BinLLM's prompt, \textit{i.e.,} only using the collaborative information. Additionally, we also study the influence of the two-step tuning by comparing a variant that employs intuitive tuning, denoted by BinLLM-IT. The comparison results are summarized in Table~\ref{tab-ablation}.

From the table, we make the following observations: 1) BinMF underperforms all BinLLM variants that consider collaborative information, confirming the superiority of leveraging LLMs for recommendation. 2) BinLLM-TO underperforms other BinLLM variants, indicating that introducing collaborative information is crucial for enhancing LLMRec performance. 3) BinLLM-IO generally underperforms BinLLM-IT and the default BinLLM, highlighting the importance of considering both textual and collaborative information. Lastly, comparing BinLLM-IT with the default BinLLM, BinLLM-IT consistently performs worse. This verifies our claims about tuning designs: directly tuning LLMs with prompts containing collaborative information from scratch may lead to underutilization of both textual and collaborative information.

\noindent \textit{The influence of text-like encoding method}.  
Taking a further step, we explore how the performance changes when the collaborative information is encoded in alternative textual formats instead of binary sequences. Specifically, we examine a variant called BinLLM-emb-str, which employs UMAP~\cite{umap} to reduce the dimensionality of the embeddings and converts the results into strings for integration into our prompt. We compare this variant with the original BinLLM and CoLLM on the ML-1M dataset, yielding the following AUC results: $0.7343$ for BinLLM-emb-str, $0.7425$ for BinLLM, and $0.7243$ for CoLLM. As the results indicate, directly converting the original embeddings into strings leads to inferior performance compared to our proposed method. However, it is noteworthy that BinLLM-emb-str still outperforms CoLLM. This finding suggests that encoding collaborative information in text formats is usually advantageous, compared to the method (CoLLM) performed in latent space.


\begin{figure}[t]
\centering
\subfigure[\textbf{ML-1M}]{\includegraphics[width=1.475in,height=1.15in]{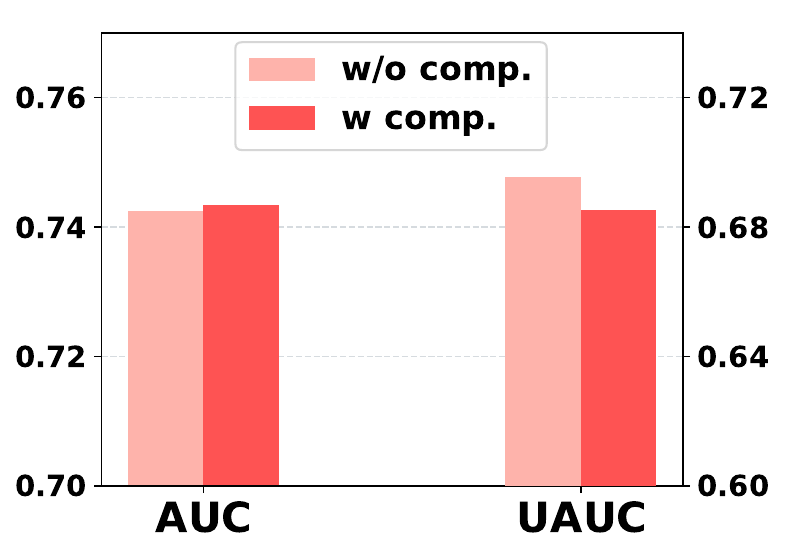}}
\subfigure[ \textbf{Amazon-book}]{ \includegraphics[width=1.475in,height=1.15in]{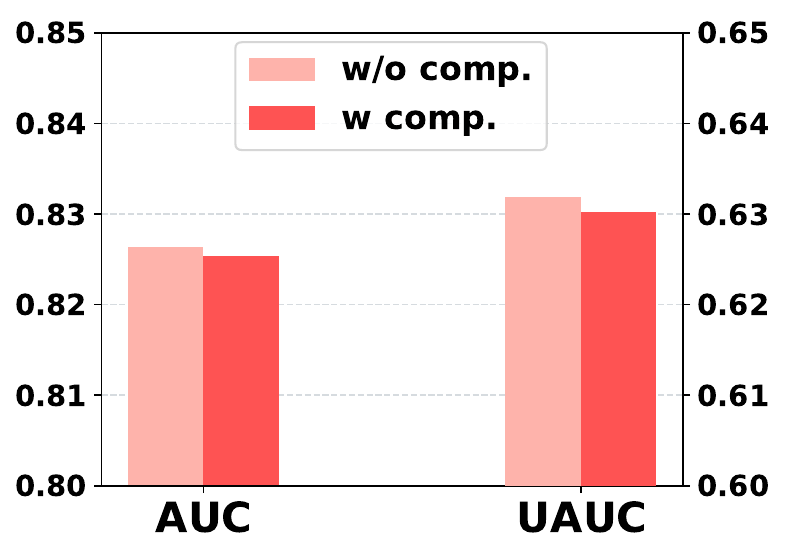}}
\caption{
Performance of BinLLM with (w comp.) and without compression (w/o comp.). The left y-axis represents AUC, while the right one represents UAUC.
}
\label{fig:compression}
\end{figure}

\subsubsection{The Influence of Compression}
In the preceding experiments, we did not use compression for our text-like encoding of collaborative information by default. Here, we conduct experiments to study its influence by comparing BinLLM with compression (w comp.) and without compression (w/o comp.). The comparison results of recommendation performance are summarized in Figure~\ref{fig:compression}. According to the figure, BinLLM with compression generally shows comparable performance to BinLLM without compression. Moreover, when compared with baselines, the comparison trends are similar to BinLLM without compression (with only some differences observed for the UAUC metric on the ML-1M dataset when compared with CoLLM). These results indicate that compression can reduce the representation length while maintaining performance to a large extent.

As shown in the example of Equation~\eqref{eq:ip}, the dot-decimal notation can compress the length of collaborative representation by approximately 2.5 times\footnote{Ignoring the dot "." in the sequence.}. However, in our experiments, the inference acceleration did not reach this level. This is because we only included the collaborative representations for the target user and items, which constitute a smaller part of the total prompt. Specifically, the inference time for BinLLM without compression and with compression was 106s and 93s on ML-1M, and 483s and 435s on Amazon, respectively. If considering collaborative information for all historically interacted items, as done by \citet{llara}, the expected inference acceleration would be more significant.

%% file: latex/5_related.tex
\section{Related Work}
$\bullet$ \textbf{Collaborative Information Modeling.} Collaborative information modeling is pivotal for personalized recommendations, and significant efforts have been dedicated to this area in traditional research. Initially, the information modeling relied on statistical methods~\citep{sarwar2001item}. Subsequently, latent factor models became prevalent, leading to the development of prominent models such as MF~\cite{koren2009matrix} and FISM~\cite{fism}. Later, neural network-enhanced latent factor models made substantial advancements~\cite{ncf, Caser, GRU4Rec,wule}. 
These studies achieved remarkable success in both academia and industry, inspiring exploration into collaborative information modeling for LLMRec. In this study, we propose a method to encode collaborative information in a text-like format, making it suitable for LLM usage.

\noindent $\bullet$ \textbf{LLMRec.} As the impressive capabilities exhibited by LLMs, an increasing number of researchers in the recommendation community are now exploring the potential of applying LLMs to recommendation systems~\cite{llmrecsurvey1, llmrecsurvey2,li2024survey}. This exploration can be categorized into two groups. The first group focuses on directly harnessing the abilities of LLMs by employing suitable prompts to stimulate their performance in recommendation scenarios~\cite{junxu_rec, hou2023large,wentao}. On the other hand, another group of researchers argues that LLMs have limited exposure to recommendation tasks during pre-training, and recommendation data often possess personalized characteristics~\cite{tallrec, instructfollows}. Consequently, it becomes crucial to explore tuning methods that can enhance the recommendation performance of LLMs~\cite{rella,textrich,lin2024data,lin2023multi}. As researchers delve deeper into their studies, it has been discovered that LLMs often exhibit an excessive reliance on semantic knowledge for learning, while paying insufficient attention to the acquisition of collaborative information between entities~\cite{bigrec}.

Researchers have initiated endeavors to incorporate collaborative information into LLMs.
Some researchers attempt to look for ID encoding methods to introduce new tokens through vocabulary expansion and train these tokens from scratch~\cite{RUCidcf, yongfengID, tiger,jundongli}. 
Among them, ~\citeauthor{yongfengID} utilize statistical information, ~\citeauthor{RUCidcf} and ~\citeauthor{tiger} employ vector quantization techniques. However, this approach often faces with low learning efficacy. Another group of researchers explores using a latent factor model to capture collaborative information~\cite{collm, e4srec, llara}, which is then mapped onto the semantic space of LLMs through a mapping layer. This method exhibits better learning efficacy but requires additional training of the mapping layer. Moreover, due to the non-text-like format of collaborative information, both sets of methods face challenges in aligning with the information processing mechanism in LLMs, limiting their performance.

%% file: latex/6_conclusion.tex
\section{Conclusion}

In this study, we emphasize the importance of text-like encoding of collaborative information modeling to enhance recommendation performance for LLMRec. We introduce BinLLM, a novel approach designed to incorporate collaborative information in a text-like format by binarizing collaborative embeddings for LLMRec. This encoding allows the collaborative information to be utilized in a manner better aligned with how information is processed in LLMs. Extensive results demonstrate the superiority of BinLLM.


%% file: latex/7_Appendenx.tex
\section*{Limitations}
Currently, this paper has certain limitations in experimental validation: 1) It relies solely on Vicuna-7B for experiments; 2) The current experiments focus solely on rating/click prediction tasks, neglecting other recommendation tasks like next-item prediction. In the future, we aim to expand experiments accordingly. Additionally, at the methodological level, similar to existing LLMRec methods, this paper faces challenges with low inference efficiency for real-world recommendation scenarios, particularly in the all-ranking setting. In the future, we could explore applying existing acceleration methods like pruning to improve speed. Moreover, exploring recommendation generation methods that avoid multiple inferences for individual users is another avenue worth exploring.

\section*{Ethical Considerations}




In this paper, we present BinLLM, designed to encode collaborative information in a text-like format for LLMRec. Our method binarizes numerical embeddings and thus doesn't raise ethical concerns. Moreover, the data we use are publicly available and don't include sensitive details like gender. However, recommendations involve user behavioral data, which might raise privacy concerns, which are addressable through introducing the mechanism of user consent. Additionally, using LLMs may have hidden negative societal biases. We advocate for conducting thorough risk assessments and advise users to be wary of potential risks linked with model usage.

\section*{Acknowledgments}
This work is supported by the National Key Research and Development Program of China (2022YFB3104701) and the National Natural Science Foundation of China (62272437).